\documentclass[twocolumn,showpacs,floatfix,superscriptaddress,amsmath,amssymb,pre]{revtex4}
\usepackage{mathrsfs}
\usepackage{txfonts}
\usepackage{amssymb}
\usepackage{graphicx}
\usepackage{hyperref}
\usepackage{ulem}
 \usepackage{overpic}

\usepackage{color}
 \usepackage{psfrag}
\usepackage{tabularx}
\usepackage{array}
\usepackage{placeins}
\makeatletter

\begin{document}

\title{ The ground-state phase diagram for an alternative anisotropic  extension of quantum spin-1 ferromagnetic biquadratic model}

\author{Yan-Wei Dai}
\affiliation{Centre for Modern Physics,
Chongqing University, Chongqing 400044, The People's Republic of
China}

\author{Qian-Qian Shi}
\affiliation{Centre for Modern Physics,
Chongqing University, Chongqing 400044, The People's Republic of
China}

\author{Xi-Hao Chen}
\affiliation{Research Institute for New Materials and Technology, Chongqing University of Arts and Sciences,
Chongqing 400000, The People's Republic of China}
\affiliation{Centre for Modern Physics,
Chongqing University, Chongqing 400044, The People's Republic of
China}

\author{Huan-Qiang Zhou}
\affiliation{Centre for Modern Physics,
Chongqing University, Chongqing 400044, The People's Republic of
China}

\begin{abstract}
The ground-state phase diagram is mapped out for an  alternative anisotropic extension of quantum spin-1 ferromagnetic  biquadratic model, which accommodates
twelve distinct phases: three degenerate fractal  phases, six Luttinger liquid phases and three symmetry-protected trivial phases.
It is found that distinct types of  quantum phase transitions are involved between them. In particular,
one type arises from an instability of a Luttinger liquid towards a degenerate fractal phase, and
the other type describes spontaneous symmetry breaking with type-B Goldstone modes from one degenerate fractal phase
to another degenerate fractal phase, with the fractal dimension $d_f$ being identical to the number of the type-B Goldstone modes, both of which turn out to be one.
In addition, quantum phase transitions from the Luttinger liquid phases to
the symmetry-protected trivial phases are identified to be in the Kosterlitz-Thouless universality
class, with central charge being one.

\end{abstract}

\maketitle
\section{Introduction}
Quantum critical phenomena continue to be a research subject of wide interest in condensed matter physics.
In the Landau-Ginzburg-Wilson (LGW) paradigm,
it is a basic notion of spontaneous symmetry breaking (SSB)~\cite{SSB} that makes it possible to classify distinct
types of quantum phase transitions (QPTs) and quantum states of matter. However, many examples, which do not fall into this paradigm, are known
even in one-dimensional quantum many-body systems, with the celebrated Haldane phase being a prominent example~\cite{haldane}.
In fact, the Haldane phase is a typical example for the so-called symmetry-protected topological (SPT) phases~\cite{wen,pollmann1}.
Further development unveils that there exist a class of the symmetry-protected trivial (SPt) phases~\cite{pollmann2}, featuring that they are
adiabatically connected to (unentangled) factorized states.

In addition, the Mermin-Wagner-Coleman theorem forbids continuous symmetries to be spontaneously broken for
quantum many-body systems in one spatial dimension~\cite{Mermin}. However, this is {\it only} valid for SSB with type-A Goldstone modes (GMs)~\cite{watanabe,nambu}.
Instead,  gapless low-lying excitations still survive strong quantum fluctuations, with their number being counted by central charge in conformal field theory~\cite{cft}.
More precisely, though (gapless) type-A GMs do not survive, their variants occur in the guise of gapless low-lying excitations in one-dimensional quantum many-body systems at criticality.

This scenario is observed in numerical simulations of one-dimensional quantum many-body systems in the context of the infinite Matrix Product State (iMPS)
representation~\cite{Wang}. In practice, the algorithms, which yield a ground-state wave function in the iMPS representation, lead to infinitely degenerate ground states in a critical regime,
due to the finiteness of the bond dimension, it thus results in pseudo SSB that vanishes as the bond dimension tends to infinity~\cite{Wang,wanghl}.
This offers a powerful means to characterize the Kosterlitz-Thouless (KT) phase transitions~\cite{KT}, which describes the instability of
a Luttinger liquid (LL) under a marginal perturbation, in the context of tensor network simulations~\cite{Vidal,McCulloch}.

In contrast, SSB with type-B GMs does occur in one spatial dimension~\cite{watanabe}. As demonstrated in Refs.~\cite{shiqq,shiqqNB,golden}, this leads to scale-invariant quantum states of matter, so an abstract fractal underlies the ground-state subspace, characterized in terms of the fractal dimension introduced by Castro-Alvaredo and
Doyon~\cite{doyon} (also cf.~\cite{popkovmt}) for the ${\rm SU}(2)$ Heisenberg ferromagnetic states. In fact,  the fractal dimension may be identified with the number of
type-B GMs. As a consequence, highly degenerate ground states arising from SSB with type-B GMs
are scale-invariant, but not conformally invariant~\cite{shiqq,shiqqNB}, thus unveiling a deep connection between scale-invariant states and the counting rule of
the GMs~\cite{watanabe, nambu}.

Therefore, it is highly desirable to search for one-dimensional quantum many-body systems that exhibit distinct types of QPTs
involving scale-invariant quantum states of matter.
In this work,  we investigate an  alternative anisotropic extension of quantum spin-1  ferromagnetic biquadratic model, both numerically and analytically. It is different from 
an anisotropic extension of quantum spin-1 ferromagnetic biquadratic model studied in Ref.~\cite{shiqq}.
Numerical simulations are carried out in terms of infinite Time Evolving Block Decimation (iTEBD)~\cite{Vidal}.
The model exhibits twelve distinct phases, accommodating three degenerate
fractal (DF) phases, six LL phases and three SPt phases.
Two novel types of QPTs are unveiled:
one type arises from the instabilities of the LL phases towards the DF phases, and
the other type describes SSB with type-B GMs from one DF phase
to another DF phase, with the fractal dimension $d_f$ being identical to the number of type-B GMs $N_B$: $d_f = N_B =1$.
In addition, QPTs from the LL phases to
the SPt phases are identified to be in the KT universality
class, with central charge $c$ being one.

    \begin{figure}
    \includegraphics[width=0.3\textwidth]{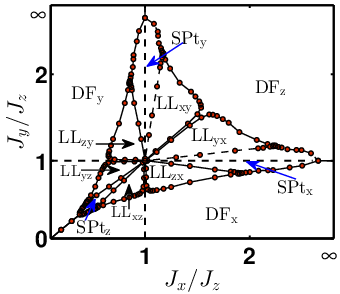}
    \caption{(color online) The ground-state phase diagram for an  alternative anisotropic extension of quantum spin-1 ferromagnetic  biquadratic model  (\ref{ham1}), which accommodates the
     DF phases, the LL phases and the SPt phases in the region
    $0 \leq J_x/J_z $, $0 \leq J_y/J_z $.
    Here, the bond dimension $\chi$ is chosen to be $\chi=200$.}
     \label{phaseD}
     \end{figure}

\section{The model and its distinct phases}
The Hamiltonian for an  alternative anisotropic extension of quantum spin-1 ferromagnetic  biquadratic model takes the form
\begin{equation}
 H = \sum_{j}
 (J_{x} S_{x, j}S_{x, j+1}+J_{y} S_{y, j}S_{y, j+1}+J_{z}S_{z, j}S_{z, j+1})^{4},
 \label{ham1}
\end{equation}
where $S_{\mu, j}\;(\mu=x,y,z)$ are the spin-1 operators at a lattice site $j$,
and $J_{\mu}$ denote the coupling parameters describing anisotropic interactions.
Here, we remark that the sum over $j$ is from 1 to $L-1$ under the open boundary conditions (OBCs) and from 1 to $L$ under the periodic boundary conditions (PBCs), with $L$ being the system size.
The model (\ref{ham1}) features one symmetric transformation and two duality transformations, as a result of its invariance under a permutation with respect to $x$, $y$ and $z$
(for more details, cf. Sec.~A  of the Supplementary Material (SM)).  That is, the symmetric and  duality transformations are induced from the symmetric group $S_3$, consisting of the permutations with respect to $x$, $y$ and $z$.  As a consequence, one may set $J_z$ as an energy scale and choose $J_x$ and $J_y$ as two independent coupling parameters.

The peculiarity of the model (\ref{ham1}) is that the symmetry group varies with $J_x$ and $J_y$.
On the characteristic line $J_x=J_y$, the model (\ref{ham1}) possesses the staggered $\rm{SU}(2)$ symmetry group  and the staggered ${\rm U(1)}$ symmetry group, generated by
$K_x$, $K_y$ and $K_z$, and $R_z$, respectively:
$K_x=\sum_jK_{x,j}$, $K_y=\sum_jK_{y,j}$ and $K_z=\sum_jK_{z,j}$, with $K_{x,j}=\sum_j(-1)^j[{S_{x,j}}^2-{S_{y,j}}^2]/2$, $K_{y,j}=\sum_j(-1)^j(S_{x,j}S_{y,j}+S_{y,j}S_{x,j})/2$ and $K_{z,j}=\sum_{j}S_{z,j}/2$,  and  $R_z=\sum _j (-1)^{j}(S_{z,j})^2$. Here, the generators $K_x$, $K_y$ and $K_z$ satisfy the commutation relations: $[K_\alpha, K_\beta]=i\varepsilon_{\alpha\beta\gamma} K_\gamma$,
where $\varepsilon_{\alpha\beta\gamma}$ is a completely antisymmetric tensor, with $\varepsilon_{xyz}=1$,
and $\alpha,\beta,\gamma = x,y,z$.  Meanwhile, a staggered $\rm{SU}(2)$ symmetry group  and a staggered ${\rm U(1)}$ symmetry group also occur on the characteristic lines $J_y=J_z$ and $J_x=J_z$, due to the symmetric and  duality transformations induced from the symmetric group $S_3$.
In particular, the staggered $\rm{SU}(3)$ symmetry group emerges at the isotropic point $J_x=J_y=J_z$. 
Given the staggered nature of the symmetry groups, we restrict ourselves to even $L$'s.
More details about the symmetry groups are described in Sec.~B of the SM.

Therefore, we may restrict ourselves to the region $J_{x}/J_{z}\geq 0$, $J_{y}/J_{z}\geq 0$ and $J_{z}=1$. In fact, the region is partitioned into six
different regimes, which are symmetric or dual to each other, as a result of the symmetric and duality transformations. Hence, we only need to focus on one of the six regimes to
perform numerical simulations of the model and map out the ground-state phase diagram.
This regime is chosen to be $0 \leq J_x\leq J_z$ and $0 \leq J_y\leq J_z$. The numerical simulations have been carried out in terms of the iTEBD~\cite{Vidal}, with the bond dimension $\chi$ being $\chi=200$. We plot
the ground-state phase diagram in Fig.~\ref{phaseD}, which accommodates the
twelve distinct phases: three DF phases, labeled as $\rm{DF_{x}}$, $\rm{DF_{y}}$ and $\rm{DF_{z}}$, six LL phases, labeled as $\rm{LL_{xy}}$, $\rm{LL_{yz}}$, $\rm{LL_{zx}}$, $\rm{LL_{yx}}$, $\rm{LL_{xz}}$ and $\rm{LL_{zy}}$,  and three SPt phases, labeled as $\rm{SPt_{x}}$, $\rm{SPt_{y}}$  and $\rm{SPt_{z}}$, respectively.
We remark that the phase boundaries are determined from the entanglement entropy~\cite{vidal} and the ground-state fidelity
per lattice site~\cite{Zhou}.
In addition, QPTs between the LL phases and the SPt phases are identified to be in the KT universality
class with central charge $c=1$, which arise from the instabilities of  the LL phases towards the SPt phases.

\begin{figure}
\includegraphics[width=0.3\textwidth]{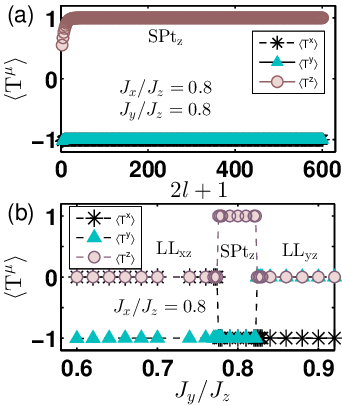}
\caption{(color online) (a) The non-local order parameter $\langle T^{\mu} \rangle$
          as a function of the block size $2l+1$ from the iTEBD simulations, with the bond dimension $\chi=60$, for an  alternative anisotropic extension of quantum spin-1 ferromagnetic biquadratic model  (\ref{ham1}), with
           $J_x/J_z= 0.8$ and $J_y/J_z=0.8$, in the $\rm {SPt_z}$ phase.
           (b) The non-local order parameter $\langle T^{\mu} \rangle$
          as a function of $J_y/J_z$ for fixed $J_x/J_z=0.8$ from the iTEBD simulations, with the bond dimension $\chi=60$,
          which may be used to detect QPT points.
 } \label{SPtFig}
\end{figure}


\section{Characterization of the distinct phases}
Now we turn to the characterization of the SPt phases, the LL phases and the DF phases.

\subsection{The SPt phases}
A local order parameter, with a non-zero value, depicts a quantum state of matter with symmetry-breaking order.
However, not all of quantum states of matter are subject to this description.
A remarkable example is the SPt phases~\cite{pollmann2}, which requires to introduce a non-local order parameter,
which is defined as the combined operation of the site-centered inversion
symmetry with the $\pi-$ rotation $R^\mu=\exp(i\pi S^\mu)$  around the $\mu$ axis $(\mu=x,y,z)$
in the spin space~\cite{pollmann2, Chen}
\begin{equation}
{T^{\mu}=\langle\psi| I_{(1, l)} . R_{(1,l)} ^\mu|\psi\rangle
   /tr(\lambda_A^2 \lambda_B^2) },
 \label{non-localorder}
\end{equation}
where $I$ is the site-centered inversion symmetry, namely $j\rightarrow -j$,
and $\lambda_A$ and $\lambda_B$ denote the Schmidt decomposition coefficients for a ground-state wave function $|\psi\rangle$, with the block size  being $2l+1$.
If the combined symmetry is retained, $\langle T^{\mu} \rangle$ must be equal to $\pm 1$. In particular, $\langle T^{\mu} \rangle=-1$ indicates
that a non-trivial SPt phase is involved.

In Fig.~\ref{SPtFig}(a), we plot the non-local order parameter $\langle T^{\mu} \rangle$ as a function of the block size $2l+1$, as a result of
the iTEBD simulation, with the bond dimension $\chi=60$, for the model  (\ref{ham1}) with
$J_{x}/J_{z}=0.8$ and $J_{y}/J_{z}=0.8$, located in the $\rm {SPt_z}$ phase.
Our numerical results show that the block size $2l+1$ should be large enough, to ensure that the non-local order parameter $\langle T^{\mu} \rangle$ is saturated.

In Fig.~\ref{SPtFig}(b), we plot the non-local order parameter $\langle T^{\mu} \rangle$  as a function of $J_y/J_z$ for fixed
$J_x/J_z=0.8$.
In the SPt phase, the non-local order parameters take the value:
$T^{x}= -1$, $T^{y}=-1$, and $T^{z} = 1$.
In the LL phase, pseudo SSB occurs. Hence,  the non-local order parameters  take the values:
$T^{x}= 0$, $T^{y}=-1$, and $T^{z} = 0$ in the $\rm LL_{xz}$ phase and
$T^{x}= -1$, $T^{y}=0$, and $T^{z} = 0$ in the $\rm LL_{yz}$ phase, respectively.
The results show that the non-local order parameters $\langle T^{\mu} \rangle$ detect the (pseudo) phase transition points:
one is located at $J_y/J_z=0.774$ on the boundary  between the $\rm {LL_{xz}}$ phase and the $\rm{SPt_z}$ phase, and the other is located at $J_y/J_z=0.882$
on the boundary between the $\rm{SPt_z}$ phase and the $\rm {LL_{yz}}$ phase, respectively.

   \begin{figure}
    \includegraphics[width=0.3\textwidth]{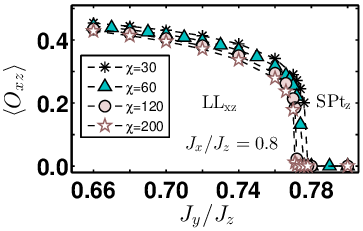}
    \caption{(color online) The pseudo order parameter $\langle O_{xz}\rangle$ as a function of
    $J_y/J_z$ for fixed $J_x/J_z=0.8$, with the bond dimension $\chi=30, 60, 120$ and $200$, respectively.
     The pseudo phase transition points between the
     $\rm {LL_{xz}}$ phase and the $\rm {SPt_z}$ phase are located from the pseudo order parameter $\langle O_{xz}\rangle$.
    } \label{PseudO}
     \end{figure}
    \begin{figure}
    \includegraphics[width=0.3\textwidth]{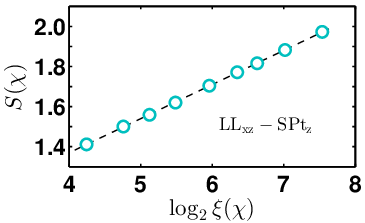}
    \caption{(color online) The entanglement entropy $S(\chi)$ versus $\log_2 \xi(\chi)$ for fixed $J_x/J_z=0.7$, at a transition point from the $\rm {LL}_{xz}$ phase to the $\rm{SPt}_z$ phase,  as a result of the iTEBD simulations. Here,
    the bond dimension $\chi$ ranges from $30$ to $200$.
    } \label{KTC}
     \end{figure}

\subsection{The LL phases}

According to the Mermin-Wagner-Coleman theorem,  continuous symmetries are not spontaneously broken for
quantum many-body systems in one spatial dimension~\cite{Mermin}. However, this is {\it only} valid for SSB with type-A GMs (GMs)~\cite{watanabe,shiqq}.
Instead,  gapless low-lying excitations still survive strong quantum fluctuations, with their number being counted by central charge $c$.
More precisely, type-A GMs survive in the guise of  gapless low-lying excitations in conformal field theories.
This scenario is observed~\cite{Wang,wanghl} in numerical simulations of one-dimensional quantum many-body systems in the context of  the iTEBD algorithm~\cite{Vidal}. In practice,
such an algorithm leads to infinitely degenerate ground states in a critical regime,
as a result of the finiteness of the bond dimension $\chi$. Hence, it results in pseudo  SSB~\cite{Wang}.
In order to keep consistency with the Mermin-Wagner-Coleman theorem, we introduce a pseudo-order parameter,
which must be scaled down to zero, when the bond dimension $\chi$ tends to infinity.

The Hamiltonian~(\ref{ham1}) possesses three $\rm{U}(1)$ symmetry groups in the entire parameter space, with only two of them being independent.
In the LL phase, it is found that, pseudo SSB occurs for one of the three $\rm{U}(1)$,
as a result of the finiteness of the bond dimension $\chi$ in the iMPS representation.
In the $\rm {LL_{xz}}$ phase, this means that a pseudo local order parameter $\langle O_{xz} \rangle=\langle S_{j}^{x}S_{j}^{z} \rangle$
emerges.
In Fig.~\ref{PseudO}, we plot the pseudo order parameter $\langle O_{xz} \rangle$ as a function of
 $J_y/J_z$ for fixed $J_x/J_z=0.8$, thus yielding a pseudo phase transition point from the $\rm LL_{xz}$ phase to the $\rm {SPt_z}$ phase, located at $J_y^c/J_z=0.776, 0.774, 0.771$ and $0.77$, with the bond dimension $\chi=30, 60, 120$ and $200$, respectively.
When the bond dimension $\chi$ tends to $\infty$, the pseudo order parameter $\langle O_{xz} \rangle$ vanishes, as it should be.

 In order to characterize a LL phase, we need to extract central charge $c$. One way to do so is to perform a finite-entanglement
 scaling analysis~\cite{Tagliacozzo}
\begin{equation}
 S(\chi)=\frac{c}{6}\log_{2} \xi(\chi)+S_0(\chi),
 \label{entropy}
\end{equation}
where $S_0(\chi)$ is an additive non-universal constant, and $\xi(\chi) \sim \chi^{\kappa}$, with $\kappa$ being a finite-entanglement scaling exponent.
The best linear fit is performed for four different points in the LL phase: $(J_x/J_z, J_y/J_z)=(0.6,0.5)$, $(0.6,0.8)$,
 $(0.7,0.8)$ and $(0.8,0.9)$, respectively, with the bond dimension $\chi$ ranging from $20$ to $200$.
 As a consequence, central charge $c$ is estimated to be $c=1$, with a relative error being less than
 $2\%$. Meanwhile,
central charge $c$ is extracted from a finite-size scaling analysis by exploiting  a finite-size MPS algorithm under the periodic boundary
conditions~\cite{Verstraete} (for more details, cf. Sec.~C of the SM).

In addition, we focus on a QPT from the $\rm{LL_{xz}}$ phase to the $\rm {SPt_z}$ phase.
The finite-entanglement scaling is performed for a pseudo critical point $J_y^c/J_z$ to extract central charge $c$.
In Fig.~\ref{KTC}, we plot the entanglement entropy $S(\chi)$ versus $\log_2 \xi(\chi)$ for fixed $J_x/J_z=0.7$,
with the bond dimension $\chi$ ranging from $30$ to $200$.
The best linear fit is exploited to estimate central charge $c=1.0206$, thus we conclude that
the phase transition~\cite{KT} from a critical phase to a gapful phase, which
describes the instabilities of the LL under a marginal perturbation, is in the KT universality class, with central charge $c=1$.

\subsection{The DF phases}

\begin{figure}
    \includegraphics[width=0.45\textwidth]{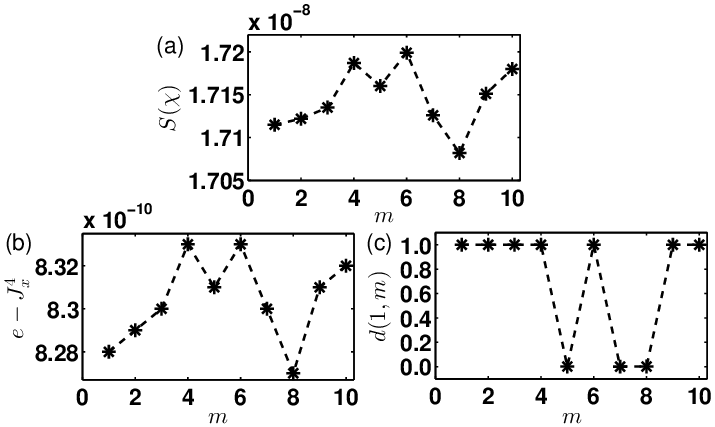}
    \caption{(color online) (a) The entanglement entropy $S(\chi)$;
                            (b) The ground-state energy per lattice site $e$, with a constant $J_{x}^4$ being subtracted;
                            and (c) the ground-state fidelity $d(1,m)$ per lattice site, with the ground state from the first
                            random trial being a reference state. We present our results from the iTEBD simulations
     as a function of  $m$, with $m$ being the number of the random trials to choose different initial states.
    Here,  $J_x/J_z=0.3$ and $J_y/J_z=0.8$ is chosen in the ${\rm DF}_y$ phase, and the
    bond dimension $\chi$ is $\chi=10$.
    } \label{DFy}
     \end{figure}

In the ${\rm DF}_y$ phase, we choose  $J_x/J_z=0.3$ and $J_y/J_z=0.8$ to perform the iTEBD simulations. Here, the bond dimension $\chi$ is chosen
to be $\chi=10$.
Hence, one may generate a sequence of the ground states $|\psi(k)\rangle$, where $k=1,2,\cdots,m$, with $m$ being the
number of the random trials to choose different initial states
In Fig.~\ref{DFy}(a), we plot the entanglement entropy $S(\chi)$ as a function of  $m$.  As we see, the entanglement entropy $S(\chi)$ is negligible, with its numerical values being
around $10^{-8}$ in magnitude. That is, no entanglement is present in the ${\rm DF}_y$ phase.
In Fig.~\ref{DFy}(b), we plot $e-J_x^4$ as a function of $m$, with its values being around $10^{-9}$ in magnitude. This implies that the ground-state
energy per lattice site $e$ is equal to $J_{x}^4$ in the $\rm{DF_y}$ phase.
In addition, we choose the ground state $|\psi(1)\rangle$ as a reference state, and
define the ground-state fidelity per lattice site $d(1,m)$ as follows~\cite{Zhou}.
The fidelity between $|\psi(1)\rangle$ and $|\psi(m)\rangle$ is  $F(|\psi(m)\rangle, |\psi(1)\rangle)=|\langle\psi(m)|\psi(1)\rangle|$,
which asymptotically scales
as $F(|\psi(m)\rangle, |\psi(1)\rangle) \sim d^L(1,m)$,
with $L$ being the system size.
In Fig.~\ref{DFy}(c), we plot the ground-state fidelity per lattice site $d(1,m)$ as a function of
$m$, which appears to take two different values $1$ and $0$. This implies
that there are two degenerate ground states in the $\rm{DF_y}$ phase.
Hence, our numerical results indicate that, in the $\rm {DF}$ phases, all the ground states are factorized states,
with the ground-state degeneracy being two.

We turn to an analytical approach to the ground states in the ${\rm DF}_y$  phase, away from the two characteristic lines $J_y/J_z=1$ with $J_x/J_z>0$  and
$J_x/J_z=0$ with $J_y/J_z>0$.
The two degenerate  ground states  in the $\rm{DF_y}$ phase, denoted as
  $|\psi_g\rangle$, take the form:
$|\psi_g\rangle=\bigotimes_l |0_y 0_z\rangle_l$ and
$|\psi_g\rangle=\bigotimes_l |0_z 0_y\rangle_l$ ($l=1,\cdots,L/2$),  where $|0_y\rangle_{2l-1/2l}$ and $|0_z\rangle_{2l-1/2l}$ are eigenvectors of the spin operators
$S_{y,2l-1/2l}$ and $S_{z,2l-1/2l}$, respectively,  with an eigenvalue being zero.
Note that the ground-state energy per lattice site $e$ being equal to $J_x^4$,  consistent with the numerical results from the iTEBD simulations.
The ground-state degeneracy is two, due to the fact that the one-site translation symmetry is spontaneously broken in the $\rm{DF_y}$ phase.

In particular, we are able to derive the ground-state wave functions on the characteristic line $J_y/J_z=1$ with $J_x/J_z>0$ (for more details,
cf. Sec.~D of the SM). Since the model possesses a staggered $\rm{SU}(2)$ symmetry group  and a staggered ${\rm U(1)}$ symmetry group  on the
characteristic line $J_y/J_z=1$ with $J_x/J_z>0$, SSB with type-B GMs arises, thus yielding highly degenerate ground states, with the number
of type-B GMs $N_B$ being one: $N_B=1$.
That is, an abstract fractal underlies the ground-state subspace, with the fractal dimension $d_f$ being identical to the number
of type-B GMs $N_B$. Combining with the symmetric and duality transformations induced from the symmetric group $S_3$, the staggered $\rm{SU}(3)$ symmetry
group emerges at the isotropic point $J_x=J_y=J_z$. In fact, the model Hamiltonian (\ref{ham1}) is unitarily equivalent to the staggered $\rm{SU}(3)$ ferromagnetic
biquadratic model, as seen in Sec.~B of the SM. Therefore,
as shown in Ref.~\cite{golden},  the staggered $\rm{SU}(3)$ symmetry
group is spontaneously broken to $\rm{U}(1) \times \rm{U}(1)$, thus yielding two type-B GMs, with the fractal dimension $d_f$ being identical to the number
of type-B GMs $N_B$: $d_f = N_B =2$.

In addition, on the characteristic line $J_x/J_z=0$ with $J_y/J_z>0$, a brute force calculation shows that a factorized
ground state $|\Psi_f\rangle$, with the ground-state energy per lattice site $e$ being equal to $0$, takes the form
\begin{eqnarray}
	|v_1\rangle_{2l-1}=&\sin\zeta|0_y\rangle_{2l-1} + e^{i\theta}\cos\zeta |0_z\rangle_{2l-1},\\ \nonumber
	|v_2\rangle_{2l}= &\frac{J_y\cos\zeta}{\sqrt{J_y^2\cos^2\zeta+J_z^2\sin^2\zeta}}|0_y\rangle_{2l} +e^{-i\theta}\frac{J_z\sin\zeta}{\sqrt{J_y^2\cos^2\zeta+J_z^2\sin^2\zeta}}|0_z\rangle_{2l},
	\label{v12}
\end{eqnarray}
where $\zeta$ and $\theta$ are two free parameters that are real
(for more details, cf. Sec. D of the SM).
That is, the model (\ref{ham1}) admits a two-parameter family of degenerate factorized ground states on the characteristic line $J_x/J_z=0$ with $J_y/J_z>0$.
Generically, $|\Psi_f\rangle$ is not invariant under the one-site translation operation, indicating that the  symmetry under the one-site translation operation is spontaneously broken.

Our discussions clearly show that the ground-state degeneracy is $L+1$ on the characteristic lines $J_x/J_z=0$ and $J_y/J_z=1$, and it is {\it only} two in the $\rm{DF_y}$ phase away from the two characteristic lines, though they are in the {\it same} $\rm{DF_{y}}$ phase. This observation challenges the conventional wisdom, which dictates that it is always possible to adiabatically connect ground-state wave functions at any two points, as long as they are in the {\it same} phase.

\section{Novel types of quantum phase transitions}

Two novel types of QPTs are unveiled. The first type arises from the instabilities of the LL phases towards the DF phases.
This type of QPTs is similar to the celebrated Pokrovsky-Talapov (PT) transitions~\cite{pt}, with a remarkable difference that the ground-state degeneracy is, generically, two in the DF phases. In contrast, the PT transitions describe the instabilities of the LL phases towards an unentangled state, with the ground-state degeneracy being one.
In both cases, no entanglement is present in a ground-state wave function that may be attributed to a (trivial) scale-invariant state, with the fractal dimension $d_f = 0$~\cite{entropy}.

The second type describes SSB with type-B GMs from one DF phase
to another DF phase, with the fractal dimension $d_f$ being identical to the number of  type-B GMs $N_B$.
We remark that the Hamiltonian~(\ref{ham1}) possesses the staggered $\rm{SU}(2)$ symmetry group, generated by $K_x$, $K_y$ and $K_z$, in addition to the $\rm{U}(1)$ symmetry group, generated by $R_z$, on the characteristic line $J_x=J_y$. We remark that the staggered $\rm{SU}(2)$ symmetry group is spontaneously broken to $\rm{U}(1)$, with  the $\rm{U}(1)$ symmetry group, generated by $R_z$, being left
intact. Accordingly, one may define the raising operator $K_+=\sum_j{K_{+,j}}$ and the lowering operator $K_-=\sum_j{K_{-,j}}$, with $K_{\pm,j}=(K_{x,j}\pm iK_{y,j})/\sqrt{2}$. They satisfy
the commutation relations: $[K_z,K_+]=K_+$, $[K_+,K_-]=K_z$ and $[K_-,K_z]=K_-$.

We choose $|{\rm hws}\rangle=(|0_x0_z...0_x0_z\rangle+|0_y0_z...0_y0_z\rangle)/\sqrt{2}$ as the highest weight state, which is invariant under the two-site translation.
Here, $|0_x\rangle/|0_y\rangle$ is the eigenvector of $S_{x,j}/S_{y,j}$, with the eigenvalue being $0$.
The interpolating fields are $K_{+,j}$ and $K_{-,j}$, for the generator $K_-$ and the generator $K_+$, respectively.
Thus, $\langle K_{z,j}\rangle$ is the local order parameter, given that $\langle[K_{+,j},K_-]\rangle=\langle[K_+,K_{-,j}]\rangle=\langle K_{z,j}\rangle\neq0$.
Therefore, the two generators $K_-$ and $K_+$ are broken.
According to the counting rule~\cite{watanabe}, the number of type-B GMs $N_B$ is one: $N_B = 1$.

As demonstrated in Ref.~\cite{shiqqNB}, for a given filling $f=M/L$, the entanglement entropy $S_f(n)$ take the form
\begin{equation}
S_f(n)=\frac{N_B}{2}\log_2 n+S_{0f},
\end{equation}
where $S_{0f}$ is an additive non-universal constant. An analytical treatment  is presented  for $S_f(n)$ in Sec.~E of the SM, confirming this scaling relation, with $N_B=1$.
Combining with a field-theoretic prediction that the prefactor is half the fractal dimension $d_f$~\cite{doyon}, we conclude that the fractal dimension $d_f$ is identical to the number of type-B GMs $N_B$: $d_f=N_B$.

\section{Summary}

An extensive numerical simulation has been performed for an  alternative anisotropic extension of quantum spin-1 ferromagnetic biquadratic model in terms of the iTEBD algorithm.
The ground-state phase diagram accommodates the
twelve distinct phases: three DF phases, six LL phases,  and three SPt phases.
In addition, an analytical approach to the DF phases has been developed to unveil a deep connection between scale-invariant states, which appear to be highly degenerate ground states arising from SSB with type-B GMs, and the counting rule of
the GMs. Meanwhile, the phase boundaries have been determined from the entanglement entropy and the ground-state fidelity
per lattice site.

As it turns out, QPTs between the LL phases and the SPt phases are identified to be in the KT universality
class with central charge $c=1$, which arise from the instabilities of  the LL phases towards the SPt phases.
In particular, two novel types of QPTs are unveiled:
one type arises from an instability of a LL phase towards a DF phase, and
the other type describes SSB with type-B GMs from one DF phase
to another DF phase, with the fractal dimension $d_f$ being identical to the number of type-B GMs $N_B$: $d_f = N_B =1$, on the characteristic line $J_x=J_y$ between the $\rm{DF_x}$ phase and $\rm{DF_y}$ phase, and  $d_f = N_B =2$ at the characteristic (isotropic) point $J_x = J_y =J_z$.
Our results challenge the conventional wisdom that it is always possible to adiabatically connect any two ground-state wave functions, as long as they are in the {\it same} phase. In fact, the ground-state degeneracies may be different for two points in the {\it same} phase, as it happens in the DF phases.

{\it Acknowledgements.}
We thank Murray Batchelor, John Fjaerestad and Ian McCulloch for enlightening discussions.


\newpage
\onecolumngrid
\newpage
\section*{Supplementary Material}
\twocolumngrid
\setcounter{page}{1}
\setcounter{equation}{0}
\setcounter{figure}{0}
\setcounter{table}{0}
\renewcommand{\theequation}{S\arabic{equation}}
\renewcommand{\thefigure}{S\arabic{figure}}
\renewcommand{\thetable}{S\arabic{table}}
\renewcommand{\bibnumfmt}{S\arabic{table}}
\renewcommand{\bibnumfmt}[1]{[S#1]}
\renewcommand{\citenumfont}[1]{S#1}

  \subsection{Symmetric and duality transformations}
\label{Duality}

We restrict ourselves to the region $J_{x}/J_{z}\geq 0$, $J_{y}/J_{z}\geq 0$ and $J_{z}=1$. This is due to the fact that the Hamiltonian~(\ref{ham1})
is invariant under permutations with respect to $x$, $y$ and $z$, which induces symmetric and duality transformations from the symmetric group $S_3$.

For simplicity, we define the variables $X=J_x/J_z$ and $Y=J_y/J_z$. Hence, the Hamiltonian~(\ref{ham1}) is re-parametrized as $H(X,Y)$.
The Hamiltonian $H(X,Y)$ satisfies the following two dualities:

(i) The Hamiltonian $H(X,Y)$ is dual to the Hamiltonian $H(X', Y')$ under a local unitary transformation
$U_{1}$:
$S_{x,j} \rightarrow - S_{x,j}$, $S_{y,j} \rightarrow S_{z,j}$ and $S_{z,j} \rightarrow S_{y,j}$.
Hence, we have $H(X,Y)=k(X,Y) U_{1} H(X',Y') U_{1}^\dag$, with
$X'=X/Y$, $Y'=1/Y$ and $k(X,Y)=1/Y^2$.
The Hamiltonian is invariant on the characteristic line $Y=1$.

(ii) The Hamiltonian $H(X,Y)$ is dual to the Hamiltonian $H(X',Y')$ under a local unitary transformation
$U_{2}$:
$S_{x,j} \rightarrow S_{z,j}$, $S_{y,j} \rightarrow -S_{y,j}$ and $S_{z,j} \rightarrow S_{x,j}$.
Hence, we have $H(X,Y)=k(X,Y) U_{2} H(X',Y') U_{2}^\dag$, with
$X'=1/X$, $Y'=Y/X$ and $k(X,Y)=X^2$.
The Hamiltonian is invariant on the characteristic line $X=1$.

In addition, the Hamiltonian $H(X, Y)$ is symmetric with respect to $X=Y$ under a local
unitary transformation $U_{0}$:
$S_{x,j} \rightarrow S_{y,j}$, $S_{y,j} \rightarrow S_{x,j}$ and $S_{z,j} \rightarrow - S_{z,j}$.

\begin{figure}[htpt]
	\includegraphics[width=0.3\textwidth]{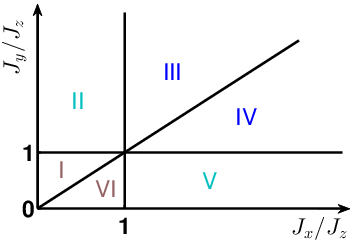}
	\caption{(color online) The six regimes,  symmetric or dual to each other, in the region $J_x/J_z\geq 0$ and $J_y/J_z\geq 0$  for an  alternative anisotropic extension of quantum spin-1 ferromagnetic biquadratic model.
	} \label{DualFig}
\end{figure}

The presence of the symmetric and duality transformations separates the region $(X\geq 0$ and $Y\geq 0)$ into
six distinct regimes, as shown in Fig.~\ref{DualFig}, which are symmetric or dual to each other.

  \subsection{ The variation of the symmetry group with the coupling parameters}
  \label{Symmetry}

 In the entire region $J_{x}/J_{z}\geq 0$ and $J_{y}/J_{z}\geq 0$, the model Hamiltonian~(\ref{ham1}) possesses three ${\rm U}(1)$ symmetry groups, generated by
  $G_{x}$, $G_{y}$ and $ G_{z}$,
 with $G_{x}= \sum _j (-1)^{j} [(S_{x,j})^2-(S_{y,j})^2]$, $G_{y}=\sum _j (-1)^{j} [(S_{y,j})^2-(S_{z,j})^2]$ and $G_{z}=\sum _j (-1)^{j} [(S_{j,z})^2-(S_{j,x})^2]$, respectively.
 However, only two of them are independent, due to the fact that $G_x+G_y+G_z=0$. Hence, the symmetry group at a generic point is ${\rm U}(1) \times {\rm U}(1)$.

The Hamiltonian~(\ref{ham1}) possesses three characteristic lines, defined as $J_x=J_y$, $J_y=J_z$ and $J_z=J_x$, respectively.
On each of the three characteristic lines, the symmetry group is enlarged to $\rm{SU}(2)\times \rm{U}(1)$.
Given that the three characteristic lines are cyclic under the symmetric group $S_3$, consisting of the permutations with respect to $x$, $y$ and $z$, one only needs to
focus on one of them. For an illustration, we choose the characteristic line $J_x=J_y$. The staggered  $\rm{SU}(2)$ symmetry group is generated by
$K_x=\sum_jK_{x,j}$, $K_y=\sum_jK_{y,j}$ and $K_z=\sum_jK_{z,j}$, with $K_{x,j}=\sum_j(-1)^j[{S_{x,j}}^2-{S_{y,j}}^2]/2$, $K_{y,j}=\sum_j(-1)^j(S_{x,j}S_{y,j}+S_{y,j}S_{x,j})/2$ and $K_{z,j}=\sum_{j}S_{z,j}/2$, and the ${\rm U(1)}$ symmetry group is generated by $R_z=\sum _j (-1)^{j}(S_{z,j})^2$.
We remark that, in three cases, the generators $K_x$, $K_y$ and $K_z$  satisfy the commutation relations: $[K_\alpha, K_\beta]=i\varepsilon_{\alpha\beta\gamma} K_\gamma$,
where $\varepsilon_{\alpha\beta\gamma}$ ($\alpha,\beta,\gamma = x,y,z$) is a completely antisymmetric tensor, with $\varepsilon_{xyz}=1$.

 In addition, the staggered $\rm{SU}(3)$ symmetry group~\cite{Dai0, shiqq0} emerges at the characteristic (isotropic) point $J_x=J_y=J_z$,
 which is realized in terms of the spin-1 operators:
 $J_{\alpha}=\sum_{j}J_{\alpha}^j$
 $(\alpha= 1, 2, \cdots, 8)$, with
 $J_1=1/2\sum_{j}S_{x,j}$, $J_2=1/2\sum_{j}S_{y,j}j^{y}$, $J_3=1/2\sum_{j}S_{z,j}$,
 $J_4=1-3/2\sum_{j}(-1)^{j}(S_{z,j})^2$, $J_5=1/2\sum_{j}(-1)^{j}({(S_{x,j})}^2-{(S_{y,j})}^2)$,
 $J_6=1/2\sum_{j}(-1)^{j}(S_{y,j}S_{z,j}+S_{z,j}S_{y,j})$, $J_7=1/2\sum_{j}(-1)^{j}(S_{z,j}S_{x,j}+S_{x,j}S_{z,j})$ and
 $J_8=1/2\sum_{j}(-1)^{j}(S_{x,j}S_{y,j}+S_{y,j}S_{x,j})$.

It appears to be proper to point out a connection of the model Hamiltonian~(\ref{ham1}) with the well-studied spin-1 ferromagnetic bilinear-biquadratic model~\cite{blbq}. In fact,
the spin-1 bilinear-biquadratic model accommodates the staggered $\rm{SU}(3)$ ferromagnetic biquadratic model, which in turn is  the isotropic limit of an alternative anisotropic  extension of quantum spin-1 ferromagnetic  biquadratic model, investigated in Ref.~\cite{shiqq0}:

\begin{equation}
	\mathscr{H} =  \sum_{j}
	( \mathscr{J}_x S_{x, j}S_{x, j+1}+ \mathscr{J}_y S_{y, j}S_{y, j+1}+\mathscr{J}_z S_{z, j}S_{z, j+1})^2.
	\label{ham2}
\end{equation}
At the isotropic point $J_x=J_y=J_z=J$ or $\mathscr{J}_x=\mathscr{J}_y=\mathscr{J}_z=\mathscr{J}$, the model Hamiltonian~(\ref{ham1}) $H$ and the Hamiltonian $\mathscr{H}$ (\ref{ham2}) satisfy the relationship:
 $H=J^4\;(5\mathscr{H}/\mathscr{J}^2-4)$, up to a unitary transformation.
As it turns out, the model Hamiltonian~(\ref{ham1}) at the isotropic point $J_x=J_y=J_z$ is unitarily equivalent to the staggered $\rm{SU}(3)$ ferromagnetic biquadratic model~(\ref{ham2}). In this sense,
one may regard the model Hamiltonian~(\ref{ham1}) as an alternative  anisotropic extension of quantum spin-1 ferromagnetic biquadratic model. Actually, the Hamiltonian ~(\ref{ham2}) shares exactly the same symmetric and  duality transformations induced from the symmetric group $S_3$ as the Hamiltonian~(\ref{ham1}). Indeed, the model (\ref{ham2}) on the characteristic lines $\mathscr{J}_x=\mathscr{J}_y$, $\mathscr{J}_y=\mathscr{J}_z$ and $\mathscr{J}_x=\mathscr{J}_z$ shares the same symmetry group as the model Hamiltonian~(\ref{ham1}) on the characteristic lines $J_x=J_y$, $J_y=J_z$ and $J_x=J_z$. Hence, both of the models in the isotropic limit share the same staggered $\rm{SU}(3)$ symmetry group. As a consequence, one may conclude that
the SSB pattern from $\rm{SU}(2)$ to
$\rm{U}(1)$ occurs  for the model Hamiltonian~(\ref{ham1}) on the characteristic line $J_y=J_z$ ($0<J_x/J_z <1$)
and on the characteristic line $J_x=J_y$ with $0<J_x/J_z<J_x^c$, with the number of type-B GMs $N_B$ being one,
and that the SSB pattern from $\rm{SU}(3)$ to
$\rm{U}(1) \times \rm{U}(1)$ occurs  for the model Hamiltonian~(\ref{ham1}) at the isotropic point $J_x=J_y=J_z$, with the number of  type-B GMs $N_B$ being two, respectively.

    \begin{figure}
    \includegraphics[width=0.46\textwidth]{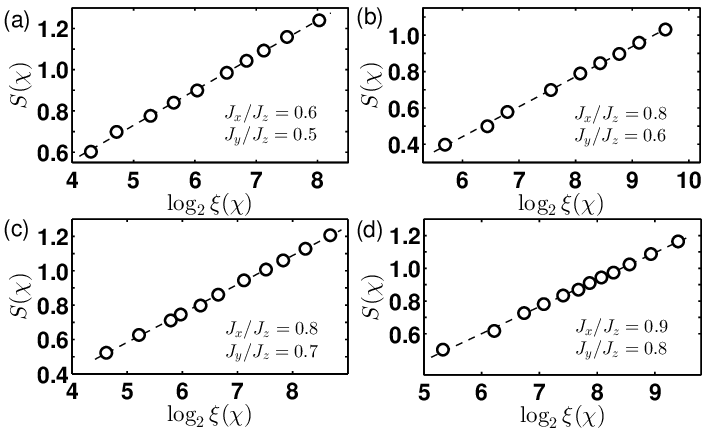}
    \caption{(color online) The entanglement entropy $S(\chi)$ versus correlation length $\xi(\chi)$ for four different points $(J_x/J_z, J_y/J_z)=(0.6,0.5)$, $(0.8,0.6)$, $(0.8,0.7)$ and $(0.9,0.8)$ in the $\rm{LL_{xz}}$ phase, with the
    bond dimension ranging from $20$ to $200$.}
    \label{CCLL}
    \end{figure}

\begin{table}
\renewcommand\arraystretch{2}
\caption{Central charge $c$ is extracted to be $c=1$ in the LL phase from the iTEBD simulations.  Here, we have chosen $(J_x/J_z, J_y/J_z)=(0.6,0.5)$, $(0.8,0.6)$, $(0.8,0.7)$ and $(0.9,0.8)$, respectively.
 }
\begin{tabular}{c|ccccccc}
\hline
\hline
      & \begin{minipage}{1.7cm} $J_x/J_z=0.6$ \\ $J_y/J_z=0.5$ \end{minipage} &
      \begin{minipage}{1.7cm} $J_x/J_z=0.8$ \\ $J_y/J_z=0.6$ \end{minipage} &
      \begin{minipage}{1.7cm} $J_x/J_z=0.8$ \\ $J_y/J_z=0.7$ \end{minipage} &
      \begin{minipage}{1.7cm} $J_x/J_z=0.9$ \\ $J_y/J_z=0.8$ \end{minipage} &
      \\
\hline
 \begin{minipage}{0.8cm}  $c$ \end{minipage}
 & 1.017 & 0.9906 & 1.0122 & 0.9906\\
 \hline
 \hline
\end{tabular}
\label{CCitebd}
\end{table}

\subsection{ Extracting central charge $c$ from the entanglement entropy in the LL phases}

Here, central charge $c$ is extracted from the iTEBD simulation~\cite{Vidal0} and from
the finite-size  MPS simulations under the PBCs~\cite{Verstraete0} in the LL phases.

\subsubsection*{1. Central charge $c$ from the {\rm iTEBD} simulations in the {\rm LL} phases}

In the iMPS representation, the entanglement entropy $S$~\cite{Bennett0}, as a measure of bipartite entanglement,
is written as $S=-\sum_{i=1}^{\chi}\lambda_{i}^{2}\log_2\lambda_{i}^{2}$,
with $\lambda_{i}$ being the singular values.
Central charge $c$ in a critical regime is estimated from a finite-entanglement scaling analysis~(\ref{entropy}).

For a given value of the  bond dimension $\chi$, the correlation length $\xi$ is defined in terms of the ratio between the
largest and second largest eigenvalues of the transfer matrix:
 $1/\xi(\chi)=\log_2|\epsilon_0(\chi)/\epsilon_1(\chi)|$.
In order to characterize the LL phases,
we choose four different points $(J_x/J_z, J_y/J_z)=(0.6,0.5)$, $(0.8,0.6)$,
 $(0.8,0.7)$ and $(0.9,0.8)$ in the $\rm {LL_{xz}}$ phase.
 The finite-entanglement scaling analysis is performed, with the bond
 dimension ranging from $20$ to $200$.
In Fig.~\ref{CCLL}, the best linear fit is exploited to estimate central charge $c$, listed in Table~\ref{CCitebd}.
The iTEBD simulations yield that central charge is $c=1$ in the $\rm {LL_{xz}}$ phase, with a relative error being less than $2\%$.

\subsubsection*{2. Central charge $c$ from the finite-size simulations in the {\rm LL} phases}

We are also able to extract central charge $c$ from numerical simulations of the model (\ref{ham1}) in the LL phases in terms of the finite-size MPS algorithm for quantum many-body systems
under the PBCs~\cite{Verstraete0}.

\begin{figure}
    \includegraphics[width=0.46\textwidth]{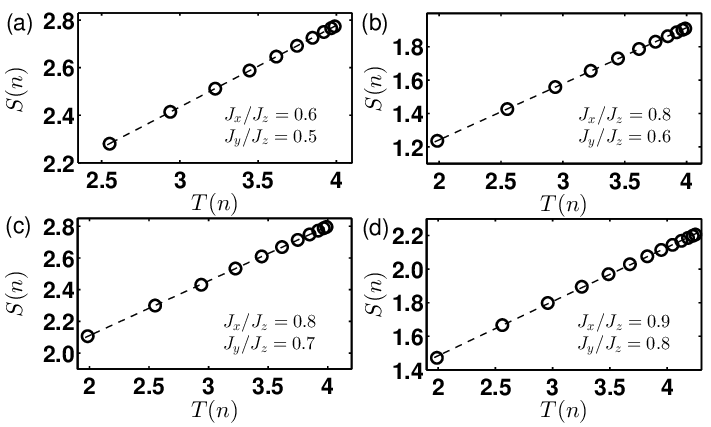}
    \caption{(color online) A finite block-size scaling analysis of the entanglement entropy $S(n)$ from the finite-size MPS simulations under the
    PBCs,
    for the four selected points $(J_x/J_z, J_y/J_z)=(0.6,0.5)$, $(0.8,0.6)$,
    $(0.8,0.7)$ and $(0.9,0.8)$
    in the $\rm{LL_{xz}}$ phase.
    Here, we have chosen the system size $L=50$, with the bond dimension $\chi=30$.}
    \label{FiniteCC}
    \end{figure}

A  quantum many-body system, with the system size being $L$, is partitioned into a block $n$ and its environment $L-n$, respectively.
Then, central charge $c$ may be extracted from a
scaling relation for the entanglement entropy $S(n)$
under the PBCs~\cite{cardy}:
\begin{equation}
{S(n)=\frac{c}{3}T(n)+S_0},
\label{entropy2}
\end{equation}
where $S_0$ is an additive non-universal constant, and $T(n)$ is a universal scaling function, defined as
\begin{equation}
{T(n)=\log_2(\frac{L}{\pi}\sin\frac{\pi n}{L})}.
 \label{to}
\end{equation}
We choose four different points  $(J_x/J_z, J_y/J_z)=(0.6,0.5)$, $(0.8,0.6)$,
$(0.8,0.7)$ and $(0.9,0.8)$ in the $\rm {LL_{xz}}$ phase, with the system size $L=50 $. The simulations are performed in terms of the finite-size MPS algorithm
under the PBCs~\cite{Verstraete0}, with the bond dimension $\chi=30$.
In Fig.~\ref{FiniteCC}, the best linear fit is performed to estimate central charge $c$, listed in Table~\ref{CCfinite}, with a relative error being less than $4\%$.
Therefore, the finite-size simulation results lend further support to our conclusion that the $\rm {LL_{xz}}$ phase is critical, with central charge being one.


\begin{table}
\renewcommand\arraystretch{2}
\caption{Central charge $c$ is extracted to be $c=1$ in the LL phase from the finite-size MPS simulations under the
	PBCs, with an error being less than $4\%$.
Here, we have chosen  $(J_x/J_z, J_y/J_z)=(0.6,0.5)$, $(0.8,0.6)$,
$(0.8,0.7)$ and $(0.9,0.8)$, with the system size $L=50 $ and the bond dimension $\chi= 30$.
 }
\begin{tabular}{c|cccccc}
\hline
\hline
      & \begin{minipage}{1.6cm} $J_x/J_z=0.6$ \\ $J_y/J_z=0.5$ \end{minipage} &
      \begin{minipage}{1.6cm} $J_x/J_z=0.8$ \\ $J_y/J_z=0.6$ \end{minipage} &
      \begin{minipage}{1.6cm} $J_x/J_z=0.8$ \\ $J_y/J_z=0.7$ \end{minipage} &
      \begin{minipage}{1.6cm} $J_x/J_z=0.9$ \\ $J_y/J_z=0.8$ \end{minipage} &
      \\
\hline
 \begin{minipage}{0.8cm}  $c$ \end{minipage}
 & 1.029 & 1.008 & 1.034 & 0.9693 & \\
 \hline
 \hline
\end{tabular}
\label{CCfinite}
\end{table}


\subsection{The factorized ground states in the $\rm{DF}$ phases: on and away from the characteristic lines}

 We emphasize that the Hamiltonian~(\ref{ham1}) is invariant under permutations with respect to $x$, $y$ and $z$, which induce symmetric and duality transformations from the symmetric group $S_3$. Hence, one may restrict to the $\rm{DF_y}$ phases. However, we need to tackle two situations separately:  on and away from the two characteristic lines:  $J_x/J_z=0$ with $J_y/J_z>0$ and $J_y/J_z=1$ with $J_x/J_z>0$.

\subsubsection*{1. The factorized ground states away from the two characteristic lines}

From our numerical simulation results, we know that the ground-state wave functions
are factorized, which are doubly degenerate in the ${\rm DF}_y$ phase, away from the two characteristic lines   $J_x/J_z=0$ with $J_y/J_z>0$ and $J_y/J_z=1$ with $J_x/J_z>0$.
Given that the Hamiltonian~(\ref{ham1}) is invariant under the one-site translation operation,
a factorized ground state $|\psi\rangle$ takes the form
$|\psi\rangle =\bigotimes_l |v_1 v_2\rangle_l$ ($l=1,\cdots,L/2$),
where
$|v_1\rangle_{2l-1} = s|0_y\rangle_{2l-1}+t |0_z\rangle_{2l-1}$
and
$|v_2\rangle_{2l}= p|0_y\rangle_{2l}+q|0_z\rangle_{2l}$, respectively, with $s$, $p$ being real numbers and $t$, $q$ being complex numbers, in order to accommodate the situation that the symmetry under the one-site translation operation is broken.
Here, the vectors $|v_1\rangle_{2l-1}$ and $|v_2\rangle_{2l}$ have been normalized. That is, we have
$s^2+t^2=1$ and $p^2+q^2=1$.

A straightforward calculation yields
\begin{equation}
h_{2l-1,2l}|v_{1}v_{2}\rangle_l = J_x^4|v_1v_2\rangle_l + V,
\label{energy1}
\end{equation}
with
\begin{equation}
\begin{aligned}
& V= \\
&[-J_xJ_y sp(J_x^2+J_y^2+3J_z^2)+J_xJ_z tq(J_x^2+3J_y^2+J_z^2)]|0_x0_x\rangle+  \\
& [sp(2J_x^2J_z^2+J_x^2J_y^2+J_y^2J_z^2+J_z^4)-J_yJ_z tq(3J_x^2+J_y^2+J_z^2)]|0_y0_y\rangle \\
& {+[tq(2J_x^2J_y^2+J_x^2J_z^2+J_y^2J_z^2+J_y^4)-J_yJ_z sp(3J_x^2+J_y^2+ J_z^2)]|0_z0_z\rangle}.
  \end{aligned}
\label{P}
\end{equation}
If $|v_1v_2\rangle_l$ is an eigenvector of $h_{2l-1,2l}$, then $V$ must vanish. Note that
the expectation value of $h_{2l-1,2l}$ takes the form
\begin{equation}
\begin{aligned}
  _l\langle v_1v_2|h_{2l-1,2l}|v_1v_2\rangle_l=J_x^4 + Q,
  \end{aligned}
\label{energy2}
\end{equation}
with
\begin{equation}
\begin{aligned}
Q=&(spJ_z-tqJ_y)^2(2J_x^2+J_y^2+J_z^2)+J_x^2(spJ_y-tqJ_z)^2.\\
  \end{aligned}
\label{energy3}
\end{equation}
Accordingly, in order to ensure that  $|v_1v_2\rangle_l$ is a ground state, we require that $V=0$. If so, we have $Q=0$. Hence, we are led to
$p^2+s^2=1$. In other words, $p^2=t^2$ and $q^2=s^2$. Substituting back into $V$, we have
$sp=0$ and $tq=0$.
This implies that either $s=0$ and $q=0$ or $t=0$ and $p=0$. That is, we have
 
\begin{equation}
\begin{aligned}
&|v_1\rangle_{2l-1} = |0_z\rangle_{2l-1}, \\
&|v_2\rangle_{2l} = |0_y\rangle_{2l}, \\
  \end{aligned}
\label{V1V2}
\end{equation}
or
\begin{equation}
\begin{aligned}
&|v_1\rangle_{2l-1}= |0_y\rangle_{2l-1}, \\
&|v_2\rangle_{2l} = |0_z\rangle_{2l}.\\
  \end{aligned}
\label{V2V1}
\end{equation}
Thus, we have derived the two factorized ground states presented in the main text:
$|\psi_g\rangle=\bigotimes_l|0_y 0_z\rangle_l$ and
$|\psi_g\rangle=\bigotimes_l |0_z 0_y\rangle_l$, with the ground-state energy per lattice site $e$ being equal to $J_x^4$.
Hence, the degenerate ground-state degeneracy is two
in the $\rm{DF_y}$ phase, consistent with our numerical simulations.

\subsubsection*{2. The factorized ground states on the two characteristic lines}
First, we focus on the characteristic line $J_x/J_z=0$ with $J_y/J_z>0$.
A straightforward calculation yields
\begin{equation}
\begin{aligned}
&\left. {h_{2l-1,2l}|v_1 v_2\rangle_l}\right.\\
&\left. {={[sp(J_y^2J_z^2+J_z^4)-J_yJ_z tq(J_y^2+J_z^2)]|0_y0_y\rangle}}\right.\\
 &\left.{+[tq(J_y^2J_z^2+J_y^4)-J_yJ_z sp(J_y^2+J_z^2)]|0_z0_z\rangle.}\right.\\
  \end{aligned}
\label{EJx0}
\end{equation}
Setting $h_{2l-1,2l}|v_1 v_2\rangle_l=0$, we have
\begin{equation}
\begin{aligned}
  {J_zsp=J_ytq}.
  \end{aligned}
\label{EJx1}
\end{equation}
Combining with the normalization conditions
$s^2+t^2=1$ and $p^2+q^2=1$, we are able to parameterize  $s, t, p$ and $q$ as follows
\begin{eqnarray}
	|v_1\rangle_{2l-1}=&\sin\zeta|0_y\rangle_{2l-1} + e^{i\theta}\cos\zeta |0_z\rangle_{2l-1},\\ \nonumber
	|v_2\rangle_{2l}= &\frac{J_y\cos\zeta}{\sqrt{J_y^2\cos^2\zeta+J_z^2\sin^2\zeta}}|0_y\rangle_{2l} +e^{-i\theta}\frac{J_z\sin\zeta}{\sqrt{J_y^2\cos^2\zeta+J_z^2\sin^2\zeta}}|0_z\rangle_{2l},
	\label{v12}
\end{eqnarray}
where $\zeta$ and $\theta$ are two free parameters that are real. We remark that
the ground-state energy per lattice site is $0$ on the characteristic line $J_x/J_z=0$, with $J_y/J_z>0$.

Next, we turn to the characteristic line $J_y/J_z=1$ with $J_x/J_z>0$. The SSB pattern from $\rm{SU}(2)\times \rm{U}(1)$ to $\rm{U}(1)\times \rm{U}(1)$ occurs, with one type-B GM.
Here, the staggered  $\rm{SU}(2)$ symmetry group is generated by the images of $K_x$, $K_y$ and $K_z$ under the cyclic permutation: $x\rightarrow y$, $y\rightarrow z$ 
and  $z\rightarrow x$. In fact, for a given representation in which one of $K_x$, $K_y$ and $K_z$ is diagonal, we have two factorized ground states that are the highest and lowest weight states, respectively, in this representation. A sequence of degenerate ground states are generated from the action of the lowering operator on the highest weight state or from the action of the raising operator on the lowest weight state. Actually, such a sequence of degenerate ground states will be explicitly constructed for the staggered $\rm{SU}(2)$ symmetry group on the characteristic line $J_x=J_y$ with $0<J_x/J_z<J_x^c$ in Sec.~E of the SM.

\subsection{The fractal dimension for highly degenerate ground states on the characteristic line $J_x=J_y$ with $0<J_x/J_z<J_x^c$}

A sequence of degenerate ground states $|L,M\rangle$ are generated from the repeated action of the lowering operators $K_{-}$ on the highest weight state $|{\rm hws}\rangle=(|0_x0_z...0_x0_z\rangle+|0_y0_z...0_y0_z\rangle)/\sqrt{2}$:
$|L,M\rangle={1}/{Z(L,M)}K_-^{\,\,\,M}|\rm{hws}\rangle$, where $Z(L,M)$ is introduced to ensure that $|L,M\rangle$ is normalized.
We remark that $|L,M\rangle$ $(M=0,...,L/2)$ span a $L+1$-dimensional irreducible representation of the staggered $\rm{SU}(2)$ symmetry group.

In order to understand SSB with one type-B GM from an entanglement perspective, the system is partitioned into a block $\mathscr{B}$ and its environment $\mathscr{E}$.
Here, the block $\mathscr{B}$ consists of $n$ lattice sites that are not necessarily contiguous, with the rest $L-n$ lattice sites constituting the environment $\mathscr{E}$.
As a convention, we demand $n\leq L/2$.
Note that $|\rm{hws}\rangle$, as an unentangled product state, is split into $|\rm{hws}\rangle_{\mathscr{B}}$ and $|\rm{hws}\rangle_{\mathscr{E}}$.
With this in mind, we introduce the counterparts of the symmetry group $\rm{SU(2)}$ in the block $\mathscr{B}$ and the environment $\mathscr{E}$, respectively.
We define the basis states $|n,k\rangle$ and $|L-n,M\rangle$ for the block $\mathscr{B}$ and the
environment $\mathscr{E}$, which take the same form as $Z(L,M)$.
Meanwhile, $Z(n,k)$ and $Z(L-n,M)$ need to be introduced to ensure that $|n,k\rangle$ and $|L-n,M\rangle$ are normalized.

For the degenerate ground states $|L,M\rangle$, we need to introduce a unit cell consisting of two nearest-neighbor sites, due to the staggered nature of the symmetry group $\rm{SU(2)}$. Therefore, there are two possible configurations: $|-1_z 0_z\rangle$
and $|1_z 0_z\rangle$ in the unit cell. Here, $|\pm 1_z\rangle$ are the eigenvectors of $S_{z,j}$, with the eigenvalues being $\pm 1$.
One may rewrite $|L,M\rangle$ as follows
\begin{align}
	|L,M\rangle=\frac{M!}{\sqrt{2^M}Z(L,M)}
	\sum_{P}(-1)^{M}|\underbrace{{-1}_z0_z...{-1}_z0_z}_{M}
	\mid\underbrace{1_z0_z...1_z0_z}_{L/2-M}\rangle,
\end{align}
where the sum $\sum_{P}$ is taken over all the permutations $P$ for a given partition $\{M, L/2-M \}$. 
This allows us to evaluate the norm $Z(L,M)$,  which takes the form 
\begin{align}
	Z(L,M)=\frac{M!}{\sqrt{2^M}}
	\sqrt{C_{L/2}^{M}}.
\end{align}
The degenerate ground states $|L,M\rangle$ admit an exact singular value decomposition:
\begin{equation}
	|L,M\rangle=\sum_{k=0}^{\min{(M,n/2)}}\lambda(L,M,k)|n,k\rangle|L-n,M-k\rangle,
	\label{svdcasei}
\end{equation}
where the singular values $\lambda(L,M,k)$ take the form,
\begin{equation}
	\lambda(L,M,k)=\sqrt{\frac{C_{n/2}^kC_{(L-n)/2}^{M-k}}{C_{L/2}^M}}.
	\label{casei}
\end{equation}
Therefore, the eigenvalues $\Lambda(L,M,k)$ of the reduced density matrix $\rho_L(n,M)$ are $\Lambda(L,M,k)=[\lambda(L,M,k)]^2$.
Hence, the entanglement entropy may be evaluated as follows
\begin{equation}
	S_L(n,M)=-\sum_k\Lambda(L,M,k)\log_2\Lambda(L,M,k).
	\label{blockS}
\end{equation}

For our purpose, we have to simplify the eigenvalues $\Lambda(L,M,k)$ by resorting to the normal distribution approximation~\cite{Allen0}
\begin{align}
C_{a}^{b}f^b(1-f)^{a-b} = \frac{1}{\sqrt{2\pi af(1-f)}} \exp[-\frac{(b-af)^2}{2af(1-f)}],
\end{align}
where $af(1-f) \gg 1$, with $f=2M/L$ being a filling factor.   Hence,
we have
\begin{align}
\Lambda(L,M,k)&=\frac{C_{n/2}^{k} f^k (1-f)^{n/2-k} C_{(L-n)/2}^{(M-k)} f^{M-k}(1-f)^{(L-n)/2-(M-k)}} {C_{L/2}^{M}f^M (1-f)^{L/2-M}}\\
 &= \frac{1}{n}\frac{1}{\sqrt{\pi \alpha}}\exp{[-\frac{(k/n-f/2)^2}{\alpha}]},
 \label{eee}
\end{align}
where $\alpha=f(1-f)(L-n)/nL$. We remark that this approach has been exploited by Popkov
and Salerno~\cite{popkov} to treat the entanglement entropy for the $\rm{SU(2)}$ spin-$1/2$ ferromagnetic states.
Substituting Eq.(\ref{eee}) into Eq.(\ref{blockS}) and replacing the sum with an integral, we obtain
\begin{align}
S_L(n,M) = \int_0^1 R \log_2 (R/n)dx,
\label{SL}
\end{align}
with
\begin{align}
R=\frac{1}{\sqrt{\pi \alpha}}\exp{[-\frac{(x-f/2)^2}{\alpha}]}.
\end{align}
For large $n$, Eq.(\ref{SL}) becomes
\begin{align}
S_L(n,M) = \frac{1}{2}\log_2[\pi ef(1-f)]+\frac{1}{2}\log_2 {[n(1-n/L)]}.
\end{align}
In the thermodynamic limit $L\rightarrow\infty$, $S_L(n,M)$ should be replaced by $S_f(n)$,
which scales with the block size $n$ as follows
\begin{align}
S_f(n) = \frac{1}{2}\log_2[\pi ef(1-f)]+\frac{1}{2}\log_2 n.
\end{align}
This confirms that the fractal dimension $d_f$~\cite{Doyon0} is identical to the number of type-B GMs $N_B$: $d_f=N_B=1$ for highly degenerate ground states on the characteristic line $J_x/J_z=J_y/J_z$ with $0<J_x/J_z<J_x^c$. Here, we remark that $J_x^c \approx 0.377 $, as follows from the iTEBD simulations.

The same treatment also works for  highly degenerate ground states, as a result of SSB from  the staggered $\rm{SU}(2)$ symmetry group to the residual symmetry group $\rm{U}(1)$ on the characteristic line $J_y=J_z$ with $0<J_x/J_z<1$.

{\it Acknowledgements.}
We thank Murray Batchelor, John Fjaerestad and Ian McCulloch for enlightening discussions.


\begin{thebibliography}{99}
\bibitem{SSB}
P. W. Anderson, Basic Notions Condensed Matter Physics,
Addison-Wesley: The Advanced Book Program (Addison-Wesley, Reading. MA, 1997).

\bibitem{haldane} F. D. M. Haldane, Phys. Lett. A \textbf{93}, 464 (1983);
F. D. M. Haldane, Phys. Rev. Lett. \textbf{50}, 1153 (1983).

\bibitem{wen}			
X. Chen, Z.-C. Gu, and X.-G. Wen, Phys. Rev. B \textbf{83}, 0355107 (2011);
X. Chen, Z.-C. Gu, and X.-G. Wen, Phys. Rev. B \textbf{84}, 235128 (2011).

\bibitem{pollmann1}	F. Pollmann and A. M. Turner, Phys. Rev. B \textbf{86}, 125441 (2012);
F. Pollmann, E. Berg, A. M. Turner, and M. Oshikawa, Phys. Rev. B \textbf{85}, 075125 (2012);
F. Pollmann and A. M. Turner, Phys. Rev. B. \textbf{86}, 125441 (2012).

\bibitem{pollmann2} Y. Fuji, F. Pollmann, and M. Oshikawa, Phys. Rev. Lett. \textbf{114}, 177204 (2015).

\bibitem{Mermin}
N. D. Mermin and H. Wagner, Phys. Rev. Lett. \textbf{17}, 1133 (1966);
S. R. Coleman, Commun. Math. Phys. \textbf{31}, 259 (1973).

\bibitem{watanabe} H. Watanabe and H. Murayama, Phys. Rev. Lett. \textbf{108}, 251602 (2012); H. Watanabe and H. Murayama, Phys. Rev. X \textbf{4}, 031057 (2014).

\bibitem{nambu}
Y. Nambu, J. Stat. Phys. \textbf{115}, 7 (2004).

\bibitem{cft} P. D. Francesco, P. Mathieu, and D. S\'{e}n\'{e}chal, Conformal Field Theory (Springer, Berlin, 1997).

\bibitem{Wang}
H.-L. Wang, J.-H. Zhao, B. Li, and H.-Q. Zhou, J. Stat. Mech. L10001 (2011);
H.-L. Wang, A.-M. Chen, B. Li, and H.-Q. Zhou, J. Phys. A: Math. Theor. \textbf{45} 015306 (2012).

\bibitem{wanghl} Y.-W. Dai, B.-Q. Hu, J.-H. Zhao, and H.-Q. Zhou,  J. Phys. A: Math. Theor. \textbf{43}, 372001 (2010);
H.-L. Wang, Y.-W. Dai, B.-Q. Hu, and H.-Q. Zhou, Phys. Lett. A \textbf{375}, 4045 (2011).


\bibitem{KT}
V. L. Berezinskii, Sov. Phys. JETP \textbf{34}, 610 (1991);
J. M. Kosterlitz and D. J. Thouless, J. Phys. C \textbf{6}, 1181 (1973).

\bibitem{Vidal}
G. Vidal, Phys. Rev. Lett. \textbf{98}, 070201 (2007).

\bibitem{McCulloch} I. P. McCulloch, J. Stat. Mech. \textbf{2007}, P10014 (2007);
F. Heidrich-Meisner, I. P. McCulloch, and A. K. Kolezhuk, Phys. Rev. B \textbf{81}, 179902 (2010).



\bibitem{shiqq}
Q.-Q. Shi, Y.-W. Dai, S.-H. Li, and H.-Q. Zhou, arXiv:2204.05692 (2022).


 \bibitem{shiqqNB}
Q.-Q. Shi, Y.-W. Dai, H.-Q. Zhou, and I. P. McCulloch, arXiv: 2201.01071 (2022).

\bibitem{golden} H.-Q. Zhou, Q.-Q. Shi, I. P. McCulloch, and M. T. Batchelor, arXiv: 2302.13126 (2023).


\bibitem{doyon}
O. A. Castro-Alvaredo and B. Doyon, J. Stat. Mech., 2011(02): P02001 (2011);
O. A. Castro-Alvaredo and B. Doyon, Phys. Rev. Lett. \textbf{108}, 120401 (2012).

\bibitem{popkovmt}
V. Popkov and M. Salerno, Phys. Rev. A \textbf{71}, 012301 (2005).


\bibitem{vidal}
G. Vidal, J. I. Latorre, E. Rico, and A. Kitaev, Phys. Rev. Lett. \textbf{90}, 227902 (2003).

\bibitem{Zhou}
 H.-Q. Zhou and J. P. Barjaktarevi\v{c}, J. Phys. A: Math. Theor. \textbf{41} 412001 (2008);
 H.-Q. Zhou, R. Or\'us, and G. Vidal, Phys. Rev. Lett. \textbf{100}, 080601 (2008).


\bibitem{Chen}
X.-H. Chen, I. P. McCulloch, M. T. Batchelor, and H.-Q. Zhou, Phys. Rev. B. \textbf{102}, 085146 (2020).


\bibitem{Tagliacozzo}
 L. Tagliacozzo, T. R. de Oliveira, S. Iblisdir, and J. I. Latorre, Phys. Rev. B \textbf{78}, 024410 (2008);
F. Pollmann, S. Mukerjee, A. M. Turner, and J. E. Moore, Phys. Rev. Lett. \textbf{102}, 255701 (2009).


\bibitem{Verstraete}
 F. Verstraete, D. Porras, and J. I. Cirac, Phys. Rev. Lett. \textbf{93}, 227205 (2004).



 \bibitem{pt} V. L. Pokrovsky and A. L. Talapov, Phys. Rev. Lett. \textbf{42}, 65 (1979).

 \bibitem{entropy} H.-Q. Zhou, Q.-Q. Shi, and Y.-W. Dai, Entropy, \textbf{24}, 1306 (2022).






\end{thebibliography}

\begin{thebibliography}{99}
\bibitem{Dai0}
 Y.-W. Dai, Q.-Q. Shi, H.-Q. Zhou, and I. P. McCulloch, arXiv:2201.01434 (2022).

\bibitem{shiqq0}
 Q.-Q. Shi, Y.-W. Dai, S.-H. Li, and H.-Q. Zhou, arXiv:2204.05692 (2022).

 \bibitem{blbq}
 G. F\'{a}th and J. S\'{o}lyom, Phys. Rev. B \textbf{44}, 11836 (1991);
 B. Sutherland, Phys. Rev. B \textbf{12}, 3795 (1975);
 M. N. Barber and M. T. Batchelor, Phys. Rev. B \textbf{40}, 4621 (1989); C. D. Batista, G. Ortiz and J. E. Gubernatis, Phys. Rev. B \textbf{65}, 180402(R) (2002); I. J. Affleck, J. Phys. Condens. Matter \textbf{2}, 405 (1990); L. Takhtajan, Phys. Lett. A \textbf{87}, 479 (1982); H. Babujian, Nucl. Phys. B \textbf{215}, 317 (1983); M. T. Batchelor and M. N. Barber, J. Phys. A. Math. Gen. \textbf{23}, L15 (1990);
 B. Aufgebauer and A. Kluemper, J. Stat. Mech. \textbf{2010}, P05018 (2010);
 R. Lundgren, J. Blair, P. Laurell, N. Regnault, G. A. Fiete, M. Greiter, and R. Thomale, Phys. Rev. B \textbf{94}, 081112 (2016);
 R. Thomale, S. Rachel, B. A. Bernevig, and D. P. Arovas,  J. Stat. Mech. \textbf{2015}, P07017 (2015).

\bibitem{Vidal0}
G. Vidal, Phys. Rev. Lett. \textbf{98}, 070201 (2007).

\bibitem{Verstraete0}
 F. Verstraete, D. Porras, and J. I. Cirac, Phys. Rev. Lett. \textbf{93}, 227205 (2004).

\bibitem{Bennett0}
C. H. Bennett, H. J. Bernstein, S. Popescu, and B. Schumacher, Phys. Rev. A \textbf{53}, 2046 (1996).

\bibitem{cardy} P. Calabrese and J. Cardy, J. Stat. Mech. P06002 (2004).


\bibitem{Allen0}
A. O. Allen, \textit{Statistics and Queueing Theory }(Academic Press, Inc., New York, 1990), Chap. 3, p. 160.

\bibitem{popkov}
V. Popkov and M. Salerno, Phys. Rev. A \textbf{71}, 012301 (2005).

\bibitem{Doyon0}
O. A. Castro-Alvaredo and B. Doyon, Phys. Rev. Lett. \textbf{108}, 120401 (2012).

\end{thebibliography}
\end{document}